\documentclass[11pt,twoside]{article}

\usepackage{ihep,epsfig,amsfonts,amssymb,amsbsy,colordvi}

\usepackage{graphicx}
\usepackage{amsmath,amssymb}
\usepackage[numbers]{natbib}

\begin{document}

\title{\bf \uppercase{Observational Features of Black Holes}}

\author{
{\bf A.F.~Zakharov$^{1,2}$\thanks{E-mail: zakharov@itep.ru},
 A.A.~Nucita$^{3}$, F.~DePaolis$^{3}$, G.~Ingrosso$^{3}$} \\ [3mm]
 \vspace*{2mm}
{\it $^1$Institute of Theoretical and Experimental Physics,
Moscow, Russia.}\\ \vspace*{1mm}
{\it $^2$Space Research Centre of Lebedev Physics Institute, Moscow.}\\ 
{\it $^3$Dipartimento di Fisica Universita di Lecce and INFN,
Sezione di Lecce,  Italy.} 
}
\date{}
\maketitle

\begin{abstract}
\par
     Recently \citet{Holz02} considered a very attracting possibility
to detect retro-MACHOs, i.e. retro-images of the Sun by a
Schwarzschild black hole. In this paper we discuss glories
(mirages) formed near rapidly rotating Kerr black hole horizons
and propose a procedure to measure masses and rotation parameters
analyzing these forms of mirages. In some sense that is a
manifestation of gravitational lens effect in the strong
gravitational field near black hole horizon and a generalization
of the retro-gravitational lens phenomenon. 
We analyze the case of
a Kerr black hole rotating at arbitrary speed for some selected
positions of a distant observer with respect to the equatorial
plane of a Kerr black hole. We discuss glories (mirages) formed
near rapidly rotating Kerr black hole horizons and propose a
procedure to measure  masses and rotation parameters analyzing
these forms of mirages. Some time ago \citet{Falcke00} suggested to
search shadows at the Galactic Center. In this paper we present
the boundaries for shadows calculated numerically. We also propose
to use future radio interferometer RADIOASTRON facilities to
measure shapes of mirages (glories) and to evaluate the black hole
spin as a function of the position angle of a distant observer.

\end{abstract}

\section{Introduction}


More than ten years ago it was predicted that profiles of lines
emitted by AGNs and X-ray binary systems\footnote{Some of them are
microquasars (for details see, for example, \cite{Grein99, Mira00,
Mira02a}).} could have an asymmetric double-peaked shape (e.g.
\cite{Chen89,Fabian89,MPS93}).
Generation of the  broad $K_\alpha$ fluorescence lines as a result
of irradiation of a cold accretion disk was discussed by many
authors (see, for example,
\cite{MPPS92,MPPS92a,Matt92,MFR93,Bao93,Mart02} and references
therein).
\citet{Popov01,Popov02} discussed influence of microlensing on the
distortion of spectral lines including Fe $K_\alpha$ line, that
can be significant in some cases. \citet{ZPJ03,ZPJ04,ZPJ04a}
showed that the optical depth for microlensing could be
significant for cosmological distributions of microlenses.
    Recent X-ray observations of Seyfert galaxies, microquasars
and binary systems
(\cite{fabian1,tanaka1,nandra1,nandra2,malizia,sambruna,
yaqoob4,ogle1,Miller02} and references therein) confirm these
considerations in general and reveal broad emission lines in their
spectra with characteristic two-peak profiles. A comprehensive
review by \citet{Fabi00} summarizes the detailed discussion of
theoretical aspects of possible scenarios for generation of broad
iron lines in AGNs. These lines are assumed to arise in the
innermost parts of the accretion disk, where the effects of
General Relativity (GR) must be taken into account, otherwise it
appears very difficult to find a natural explanation for observed
line profile.

     Numerical simulations of the line structure
are be found in a number of papers
\cite{Bao93,Koji91,Laor91,Bao92,BHO94,bromley,Fan97,pariev2,pariev1,pariev3,
Rusz00,Ma02}. They indicate that the accretion disks in Seyfert
galaxies are usually observed at the inclination angle $\theta$
close to $30^0$ or less. This occurs because according to the
Seyfert galaxy models, an opaque dusty torque surrounds the
accretion disk which does not allow us to observe the disk at
larger inclination angles.

     However, at inclination angles $\theta > 80^0$, new
observational manifestations of GR could arise. (\citet{MPS93}
discovered such phenomenon for a Schwar\-z\-s\-child black hole,
moreover the authors predicted that their results could be
applicable  to a Kerr black hole over the range of parameters
exploited). The authors mentioned that this problem was not
analyzed in detail for a Kerr metric case and it would be
necessary to investigate this case. Below we do not use a specific
model on surface emissivity of accretion (we only assume that the
emitting region is narrow enough). But general statements (which
will be described below) can be generalized to a wide disk case
without any problem. Therefore, in this paper we check and confirm
their hypothesis for the Kerr metric case and for a
Schwar\-z\-s\-child black hole using other assumptions about
surface emissivity of accretion disks. In principle, such a
phenomenon could be observed in microquasars and X-ray binary
systems where there are neutron stars and black holes with stellar
masses.


A mathematical tool to simulate spectral line shapes and results
of simulations are published in papers \cite{zakh91,zakharov01,
zakharov1,zak_rep1,zak_rep2,Zak02pr,zak_rept,Zak_rep02_xeus,
Zak_rep02_Gamma,Zak_Rep03_Lom,Zakharov_Repin_Ch03,Zak03_Sak,Zakharov_Repin_Pom04,Zak_Rep03a,Zak_Rep03b,zak_rep03,ZKLR02}.
The approach was used in particular to simulate   spectral line
shapes. For example, \citet{ZKLR02} used this approach to simulate
the influence of a magnetic field on spectral line profiles.
Recent review one the subject one could find in papers
\cite{Zak_serbia03,Zak_SPIG04}.

     Recently \citet{Holz02} have suggested that a Schwarzschild black
hole may form retro-images (called retro-MACHOs) if it is
illuminated by the Sun. We analyze a rapidly rotating Kerr black
hole case for some selected positions of a distant observer with
respect to the equatorial plane of the Kerr black hole. We discuss
glory (mirage) formed near a rapidly rotating Kerr black hole
horizon and propose a procedure to measure the mass and the
rotation parameter analyzing the  mirage shapes. Since a source
illuminating the black hole surroundings may be located in an
arbitrary direction with respect to the observer line of sight, a
generalization of the retro-gravitational lens idea suggested by
\citet{Holz02} is needed. A strong gravitational field
approximation for a gravitational lens model was considered
recently in several papers
\citep{Frittelli00,Virbhadra00,Virbhadra02,Ciufolini02,Ciufolini03,
Bozza02,Bozza04,Bozza04b,Eiroa02,Eiroa04,Sereno03,Sereno04}. However, if
we consider the standard geometry for a gravitational lens model,
namely if a gravitational lens is located between a source and
observer, then the probability to have evidences for strong
gravitational field effects is quite small, because the
probability is about $P \sim \tau_{GL} \times {R_G}/{D_S}$ where,
$\tau_{GL}$ is the optical depth for gravitational lensing and the
factor ${R_G}/{D_S}$ corresponds to a probability to have a
manifestations for strong gravitational field effects (${R_G}$ is
the Schwarzschild radius for a gravitational lens, $D_S$ is a
distance between an observer and gravitational lens).  Therefore,
the factor $R_G/D_S$ is quite small for typical astronomical
cases. However, these arguments cannot be used for the cases of a
a source located nearby a black hole.

First, it is necessary to explain differences of a considered
geometry, standard geometry of gravitational lensing (when a
gravitational lens is located roughly speaking between a source
and an observer) and a model introduced by \citet{Holz02} when an
observer is located between a source (Sun) and a gravitational
lens that is a black hole. In this paper we will consider images
formed by retro-photons, but in contrast to \citet{Holz02} we will
analyze forms of images near black holes but not a light curve of
an image formed near black hole as \citet{Holz02} did. In our
consideration a location of source could be arbitrary in great
part (in accordance with a geometry different parts of images
could be formed) \footnote{However, if a source is located between
black hole and an observer, images formed by retro-photons and
located near black holes could be non-detectable.}, for example,
accretion flows (disks) could be sources forming such images.
Since in such cases images formed by retro-photons are considered,
we call it like retro gravitational lensing even if a source is
located near a gravitational lens (a black hole) in contrast to a
standard gravitational lens model.

The plan of the paper is as follows.  In section $2$ we discuss
possible mirage shapes.
     In section~$3$ we analyze the most simple case  of
source and  observer in the black hole equatorial plane. In
section~$4$ we consider the case  of  observer at the rotation
axis of a spinning black hole. In section $5$ we consider observer
in a general position with respect to the rotation axis. The basic
characteristics of future space based radio interferometer
RADIOASTRON were given in  $6$. In section~$7$ we discuss Sgr
A$^*$ as a possible target for RADIOASTRON observations to
possibly observe such images near the black hole located in the
object. In section $8$ we discuss our results of calculations and
in section $9$ we present our conclusions.

\section{Mirage shapes}

As usual, we  use  geometrical units with $G=c=1$. It is
convenient also to measure all distances in black hole masses, so
we may set $M=1$ ($M$ is a black hole mass).
 Calculations of mirage forms are based on
qualitative analysis of different types of photon geodesics in a
Kerr metric (for references
see~\cite{Young76,chandra,zakh86,zakh89}). In fact,  we know that
impact parameters of photons are very close to the critical ones
(which correspond to parabolic orbits). One can find some samples
of photon trajectories in~\cite{zakh91,chandra}. This set
(critical curve) of impact parameters separates escape and plunge
orbits (see for details, \cite{Young76,chandra,zakh86,zakh89}) or
otherwise the critical curve separates scatter and capture regions
for unbounded photon trajectories). Therefore the mirage shapes
almost look like to critical curves but are just reflected with
respect to z-axis. We assume that mirages of all orders almost
coincide and form only one quasi-ring from the point of view of
the observer. We know that the impact parameter corresponding to
the $\pi$ deflection is close to that corresponding to a $n\pi$
deflections (n is an odd number). For more details see
\cite{Holz02} (astronomical applications of this idea was
discussed by \citet{DePaolis03} and its generalizations for Kerr
black hole are considered by \citet{DePaolis04}). We use prefix
``quasi'' since we consider a Kerr black hole case, so that mirage
shapes are not circular rings but Kerr ones. Moreover, the side
which is formed by co-moving (or co-rotating) photons is much
brighter than the opposite side since rotation of a black hole
squeeze deviations between geodesics because of Lense--Thirring
effect. Otherwise, rotation stretches deviations between geodesics
for counter-moving photons.

This assumption is based  on estimates of differences of impact
parameters for critical values and impact parameters corresponding
to $\pi$ deflection. Actually, following \cite{Holz02} we use the
approximation $b=b_c+b_d\,e^{-\Theta}$ for the impact parameter in
the case of a Schwarzschild black hole, where $\Theta$ is the
deflection angle. For  $\pi$ deflection we will take into account
the fact that it is possible to neglect the second term $b_d
e^{-\Theta} \approx 3.4823\, e^{-\Theta} \approx 0.15 $, in
comparison with the first one $b= b_c=3\sqrt{3}\approx 5.1962$.
The precision of such procedure (when we change impact parameter
corresponding to $\pi$ rotation by the critical impact parameter)
is better than 3\% (a precision of such procedure for co-moving
(co-rotating) photons propagating near a rapidly spinning black
hole is much higher). In spite of the fact that there are a number
of other corrections to improve \citet{Luminet79} approximation
(see, for example \cite{Bel02,Bozza02,Mutka02}), the black hole
spin effect gives a more important correction with respect to the
improvement which can be obtained by improving \citet{Luminet79}
approximation. For example, if we want to consider large
deflection angles, taking into account the influence of the black
hole spin on impact parameters we conclude that the spin changes
the result of a factor larger than  2.5 (since we have an impact
parameter larger than 5.19 for a Schwarzschild metric and about 2
for the extreme Kerr metric). Note that  corrections to
\citet{Luminet79} approximation can give only an improvement less
than one percent.

We would like to mention that in connection with rapidly rotating
neutron stars, some  corrections to the Schwarzschild metric are
necessary in order to analyze photon geodesic trajectories
\cite{Bel02}. However,  this procedure is not self-consistent
since a rapidly rotating neutron star may have a rather high value
of its ellipticity $e=1-(a/b)$ \citep{Salg94a,Salg94b} (i.e.
difference between equatorial and polar axes). Indeed, even Kerr
metric is not the exact approximation near rapidly rotating
neutron star surface. In this case, like for black holes, rotation
gives a more important corrections with respect to some
modifications of \cite{Luminet79} approximation.

The full classification of geodesic types for  Kerr metric is
given in \cite{zakh86}. As it was shown in this paper, there are
three  photon geodesic types: capture, scattering and critical
curve which separates the first two sets. This classification
fully depends  only on two parameters  $\xi=L_z/E$ and
$\eta=Q/E^2$, which are known as Chandrasekhar's 
constants~\citep{chandra}. Here the Carter~constant~Q~is~given~
by~\citet{carter}
\begin{equation}
   Q = p_\theta^2 + \cos^2\theta
                    \left[a^2 \left(m^2 - E^2\right) +
{L_z^2}/{\sin^2\theta}
                    \right],     \label{eq1_1}
\end{equation}
where $E=p_t$ is the particle energy at infinity, $L_z = p_\phi$
is $z$-component of its angular momentum, $m=p_ip^i$ is the
particle mass. Therefore, since photons have $m=0$
\begin{equation}
   \eta = p_\theta^2/E^2 + \cos^2\theta
                    \left[-a^2  + \xi^2/{\sin^2\theta}
                    \right].     \label{eq1_2}
\end{equation}
The first integral for the equation of photon motion (isotropic
geodesics) for a radial coordinate in the Kerr metric is described
by the following equation  \citep{carter,chandra,zakh86,zakh91b}
\begin{eqnarray}
   \rho^4 (dr/d\lambda)^2 = R(r), \nonumber 
   \end{eqnarray}
   where
   \begin{eqnarray}
R(r)=r^4+(a^2-\xi^2-\eta)r^2+2[\eta+(\xi -a)^2]r-a^2\eta,
\label{eq1_3}
\end{eqnarray}
and $\rho^2=r^2+a^2\cos^2 \theta, \Delta=r^2-2r+a^2, a=S/M^2$. The
constants $M$ and $S$ are the black hole mass and angular
momentum, respectively. Eq.~(\ref{eq1_3}) is written in
dimensionless variables (all lengths are expressed in black hole
mass units $M$).

We will consider different types of geodesics on $r$ - coordinate
in spite of the fact that these type of geodesics were discussed
in a number of papers and books, in particular in a classical
monograph by \citet{chandra} (where the most suited analysis for
our goals was given). However, our consideration  is differed even
Chandrasekhar's analysis in the following items.

i) \citet{chandra} considered the set of critical geodesics
separating capture and scatter regions as parametric functions
$\eta(r), \eta(r)$, but not as the function $\eta(\xi)$ (as we
do). However, we believe that a direct presentation of function
$\eta(\xi)$ is much more clear and give a vivid illustration of
different types of motion. Moreover, one could obtain directly
form of mirages from the function $\eta(\xi)$ (as it will be
explained below).

ii) \citet{chandra} considered the function $\eta(r)$ also for
$\eta <0$ and that is not quit correct, because for $\eta < 0$
allowed constants of motion correspond only to capture (as it was
mentioned in the book by \cite{chandra}). This point will be
briefly discussed below.

If we fix a black hole spin  parameter $a$ and consider a plane
$(\xi,\eta)$ and different types of photon trajectories
corresponding to $(\xi,\eta)$, namely, a capture region, a scatter
region  and the critical curve $\eta_{\rm crit}(\xi)$ separating
the scatter and capture regions. The critical curve is a set of
$(\xi, \eta)$ where the polynomial $R(r)$ has a multiple root (a
double root for this case). Thus, the critical curve $\eta_{\rm
crit}(\xi)$ could be determined from the system
\citep{zakh86,zakh91b}
\begin{eqnarray}
    R(r)=0, \nonumber \\
\frac{\partial R}{\partial r} (r)=0, \label{eq1_3a}
\end{eqnarray}
for $\eta \geq 0, r \geq r_+=1+\sqrt{1-a^2}$, because by analyzing
of trajectories along the $\theta$ coordinate we know that for
$\eta < 0$ we have $M=\{(\xi,\eta)|\eta \geq -a^2 + 2 a
|\xi|-\xi^2, -a \leq \xi \leq a\}$ and for each point $(\xi,
\eta)\in M$ photons will be captured. If instead $\eta <0$ and
$(\xi, \eta)~\bar{\in}~M$, photons cannot have such constants of
motion, corresponding to the forbidden region  (see,
\citep{chandra,zakh86} for details).

One can therefore calculate the critical curve $\eta(\xi)$ which
separates the capture and the scattering regions
\citep{zakh86,zakh91b}. We remind that the maximal value for
$\eta_{\rm crit}(\xi)$ is equal to 27 and is reached at $\xi=-2a$.
Obviously, if $a \rightarrow 0$, the well-known critical value for
Schwarzschild black hole (with $a=0$) is obtained.

\begin{figure}[th!]
\vspace{0.5cm}
\begin{center}
\includegraphics[width=3.5cm]{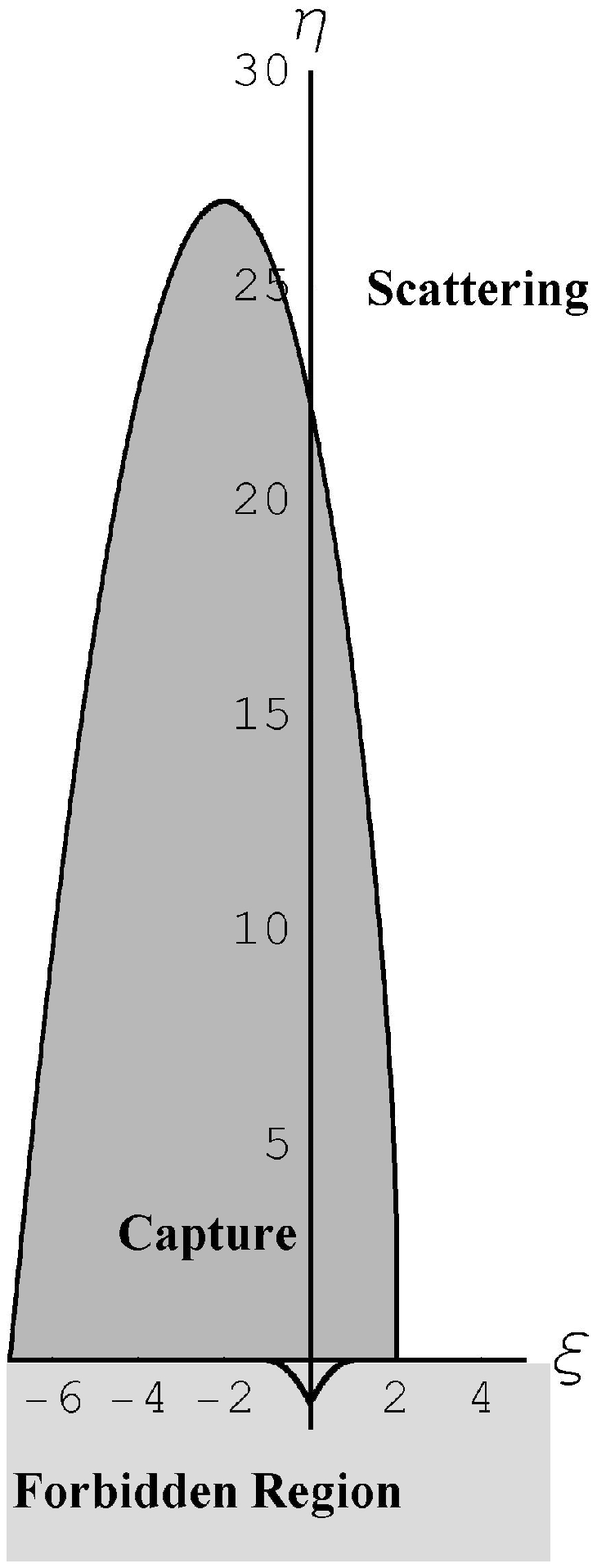}
\includegraphics[width=3.5cm]{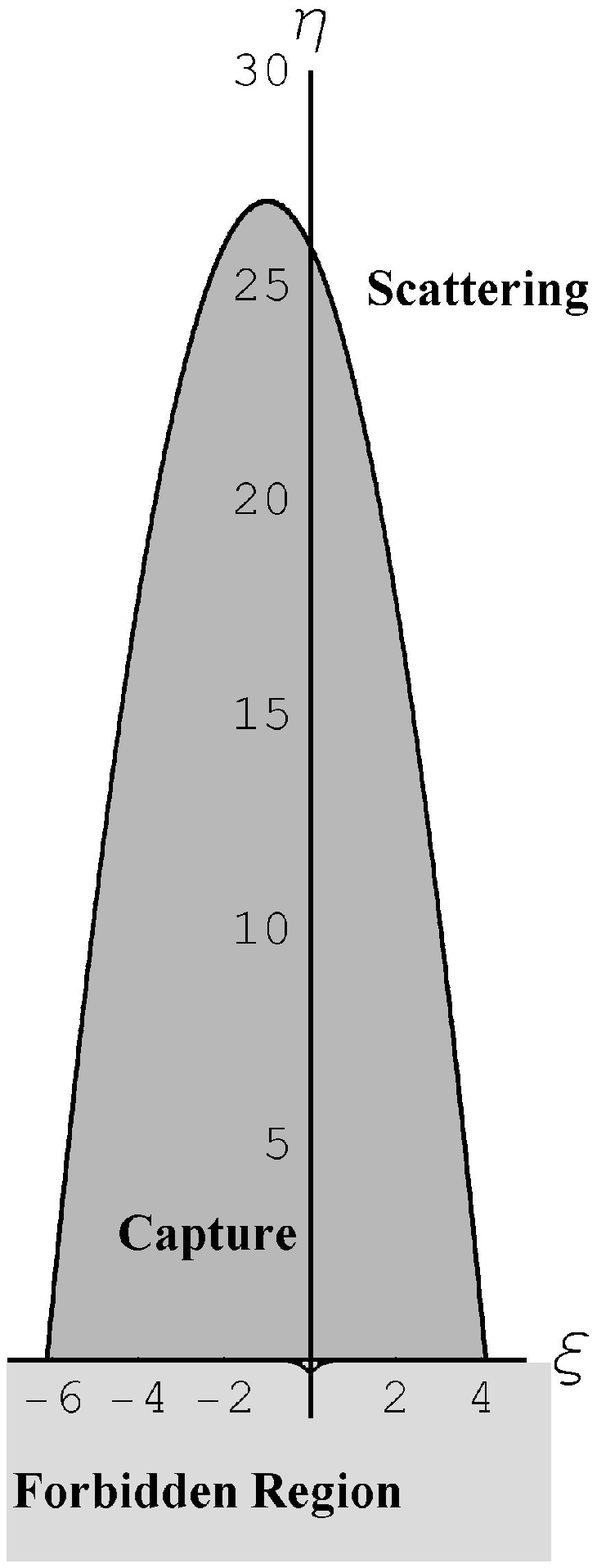}
\vspace{1cm}
\includegraphics[width=4.0147cm]{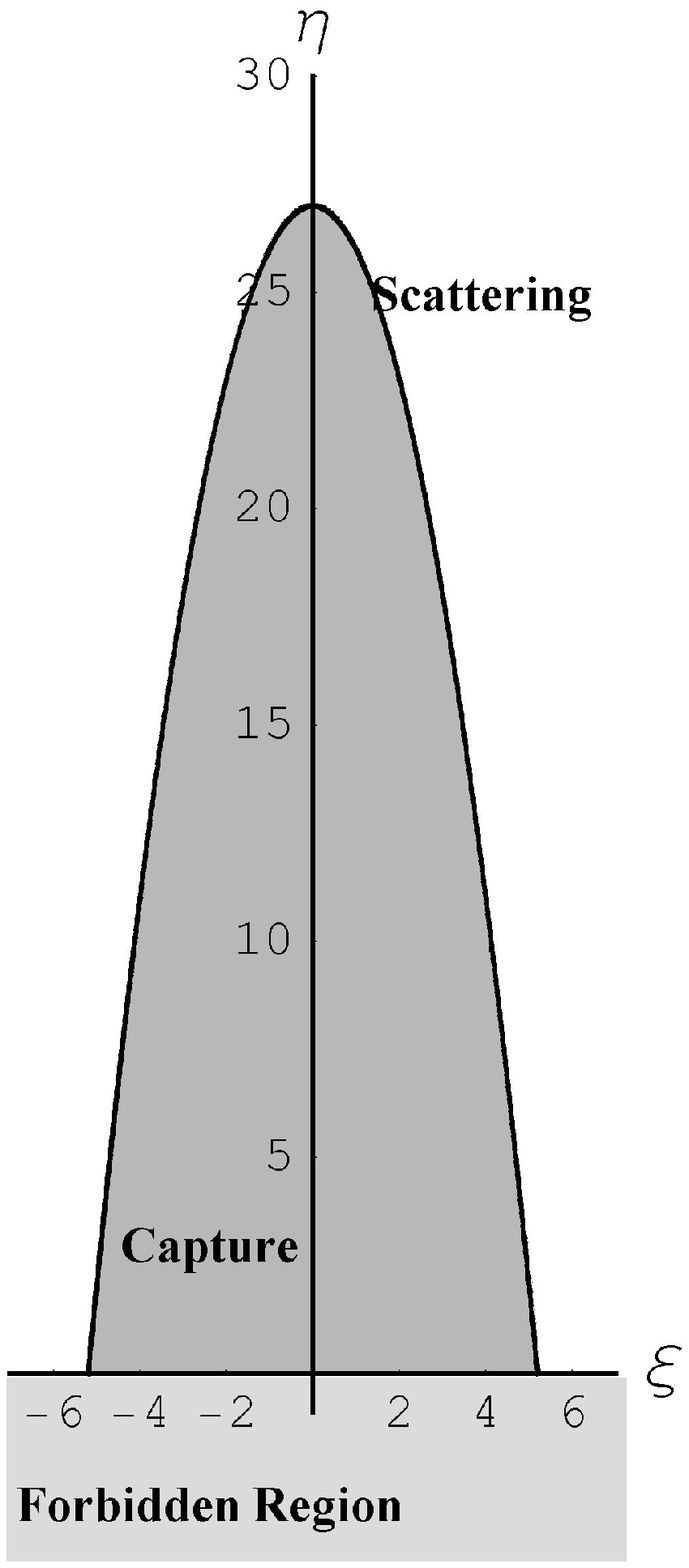}
\end{center}
\vspace*{-8mm}
\caption{Different types for photon trajectories and spin
parameters ($a=1., a=0.5, a=0.$).  Critical curves separate
capture and scatter regions. Here we show also the forbidden
region corresponding to constants of motion $\eta <0$ and $(\xi,
\eta)~\bar{\in}~M$ as it was discussed in the text.}
 \label{fig1}
\end{figure}

Thus, at first, we calculate the critical curves for chosen spin
parameters $a$ which are shown in Fig. \ref{fig1}. The shape of
the critical curve for $a=0$ (Schwarzschild black hole) is
well-known because for this case we have $\eta_{\rm crit}(\xi)=27-
\xi^2$ for $|\xi| \leqslant 3 \sqrt{3}$, but we show the critical
curve to compare with the other cases.

\vspace{0.5cm}

By following this approach  we can find the set of critical impact
parameters $(\alpha, \beta)$, for the image  (mirage or glory)
around a rotating black hole. The sets of critical parameters form
caustics around black holes and it is well-known that caustics are
the brightest part of each image (numerical simulations of caustic
formations were done by \citet{Rauch94}). We remind that $(\alpha,
\beta)$ parameters could be evaluated in terms of $(\xi, \eta_{\rm
crit})$ by the following way \citep{chandra}
\begin{eqnarray}
   \alpha(\xi)&=&\xi/\sin \theta_0,
   \label{eq1_4}
   \end{eqnarray}
  \begin{eqnarray}
\beta(\xi)&=&(\eta_{\rm crit}(\xi)+a^2 \cos^2 \theta_0 -\xi^2
\cot^2
\theta_0)^{1/2} \nonumber \\
&=& (\eta_{\rm crit}(\xi)+(a^2-\alpha^2(\xi)) \cos^2
\theta_0)^{1/2}. \label{eq1_5}
\end{eqnarray}
Actually, the mirage shapes are boundaries  for shadows considered
by \citet{Falcke00}. Recently, \citet{Beckwith04} calculated the
shapes of mirages around black holes assuming that an accretion
disk illuminates a Schwarzschild black hole.

Before closing this section, we note that the precision we obtain
by considering critical impact parameters instead of their exact
values for photon trajectories reaching the observer is good
enough. In particular, co-rotating photons form much brighter part
of images with respect to retrograde photons. Of course, the
larger is the black hole spin parameter the larger is this effect
(i.e. the co-rotating part of the images become closest to the
black hole horizon and brighter).

This approximation is based not only on numerical simulation
results of photon propagation \cite{zakharov1,zakharov5,
zak_rep1,zak_rep2,zak_rep02a,zak_rep03,zak_rep03b,zak_rep03c,zak_rep03d}
(about $10^9$ photon trajectories were analyzed) but also on
analytical results (see, for example \cite{chandra,zakh86}).

\section{Equatorial plane observer case}

\begin{figure}[th!]
\begin{center}
\includegraphics[width=10.5cm]{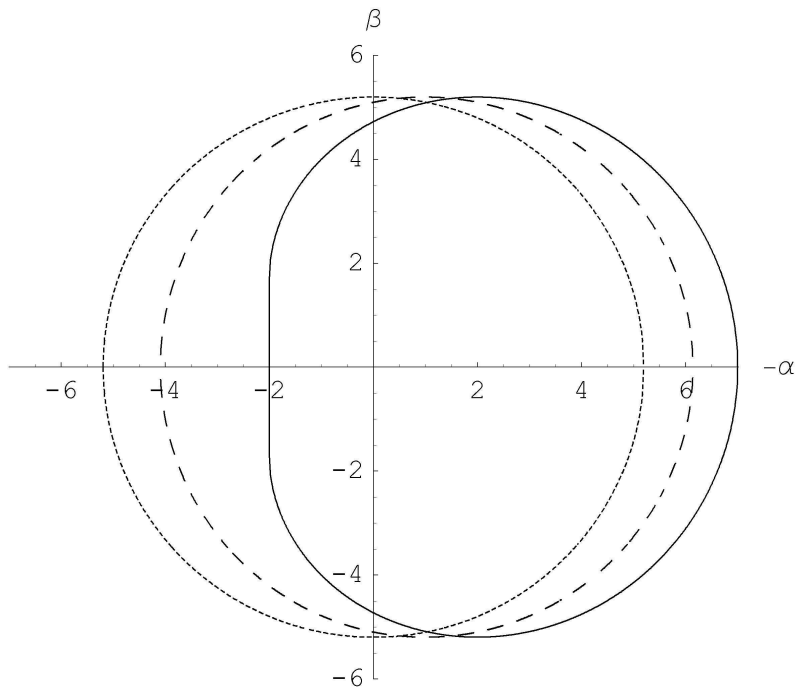}
\caption{Mirages around black hole for equatorial position of
distant observer and different spin parameters. The solid line,
the dashed line and the dotted line correspond to $a = 1, a = 0.5,
a = 0$ correspondingly.}
\end{center}
 \label{Fig2}
\end{figure}

Let us assume that the observer is located in the equatorial plane
($\theta=\pi/2$). For this case we have from Eqs.~(\ref{eq1_4})
and (\ref{eq1_5})
\begin{eqnarray}
   \alpha(\xi)&=&\xi,\\
   \label{eq2_1}
\beta(\xi)&=&\sqrt{\eta_{\rm crit}(\xi)}. \label{eq2_2}
\end{eqnarray}
As mentioned in section 2, the maximum impact value
$\beta=3\sqrt{3}$ corresponds to $\alpha=-2a$ and if we consider
the extreme spin parameter $a=1$ a segment of straight line
$\alpha=2, 0 <|\beta| < \sqrt{3}$ belongs to the mirage (see
images in Fig. \ref{Fig2} for different spin parameters). It is
clear that for this case one could easy evaluate the black hole
spin parameter after the mirage shape reconstruction since we have
a rather strong dependence of the shapes on spins. As it was
explained earlier, the maximum absolute value for
$|\beta|=\sqrt{27} \approx 5.196$ corresponds to $\alpha=-2a$
since the maximum value for $\eta(\xi)$ corresponds to
$\eta(-2a)=27$ as it was found by \citet{zakh86}. Therefore, in
principle it is possible to estimate the black hole spin parameter
by measuring the position of the maximum value for $\beta$,  but
probably that part of the mirage could be too faint to be
detected.

\section{Polar axis observer case}

If the observer is located along the polar axis we have
$\theta_0=0$ and from Eq.~(\ref{eq1_5}) we obtain
  \begin{eqnarray}
\beta(\alpha)=(\eta_{\rm crit}(0)+a^2 -\alpha^2(\xi))^{1/2},
\label{eq6_1}
\end{eqnarray}
or
\begin{eqnarray}
\beta^2(\alpha)+\alpha^2=\eta_{\rm crit}(0)+a^2. \label{eq6_2}
\end{eqnarray}
Thus, mirages around Kerr black hole look like circles and even
for this case in principle we could evaluate the black hole spin
(if the black hole mass is known) taking into account that radii
of these circles weakly depend on the black hole spin parameter.
However, one should mention that due to the small difference
between radii for different spins, even in the future it is
unlikely to be able to measure black hole spins in this way (see
Table 1).

\begin{figure}[h!]
\begin{center}
\includegraphics[width=10.5cm]{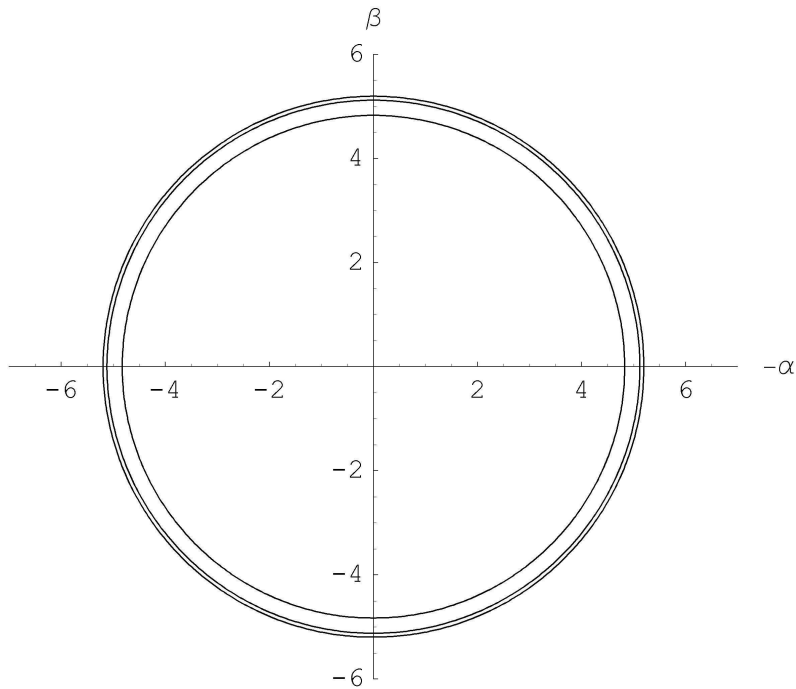}
\end{center}
\caption{Mirages around a black hole for the polar axis position
of distant observer and different spin parameters ($a=0, a=0.5,
a=1$). Smaller radii correspond to greater spin parameters.}
 \label{Fig2_1}
\end{figure}

\begin{table}[th!]
\begin{center}
\caption[]{Dependence of $\eta(0)$ and  mirage radii $R_{\rm
circ}= (\eta(0)+a^2)^{1/2}$ on spins.}
\begin{tabular}{|c|c|c|c|c|c|c|c|}
\hline
$ a$ & 0 & 0.2 & 0.4 &0.5 & 0.6 & 0.8 & 1. \\
\hline \hline
$\eta(0)$ & 27 & 26.839 & 26.348 & 25.970& 25.495& 24.210 & 22.314\\
\hline
$R_{\rm circ}$& 5.196 & 5.185 & 5.14
9  & 5.121  & 5.085 & 4.985 & 4.828 \\
\hline
\end{tabular}
\end{center}
\label{tabl0}
\end{table}

\section{General case for the angular position of the observer}

\begin{figure}[th!]
\begin{center}
\includegraphics[width=10.5cm]{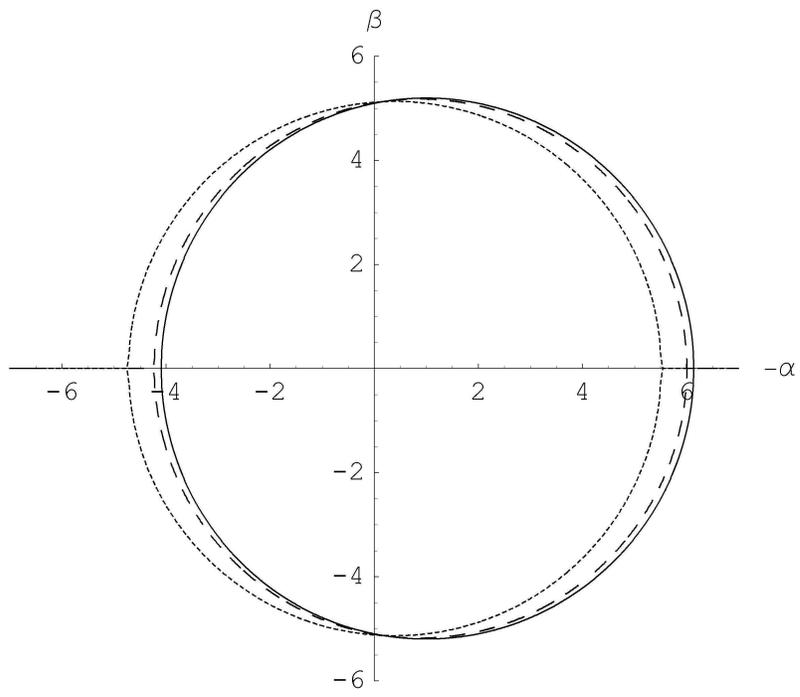}
\end{center}
\caption{Mirages around black hole for different angular positions
 of a distant observer and  the spin $a=0.5$. Solid,
dashed and dotted lines correspond to $\theta_0=\pi/2, \pi/3$ and
$\pi/8$, respectively.}
 \label{Fig2a}
\end{figure}

\begin{figure}[th!]
\begin{center}
\includegraphics[width=10.5cm]{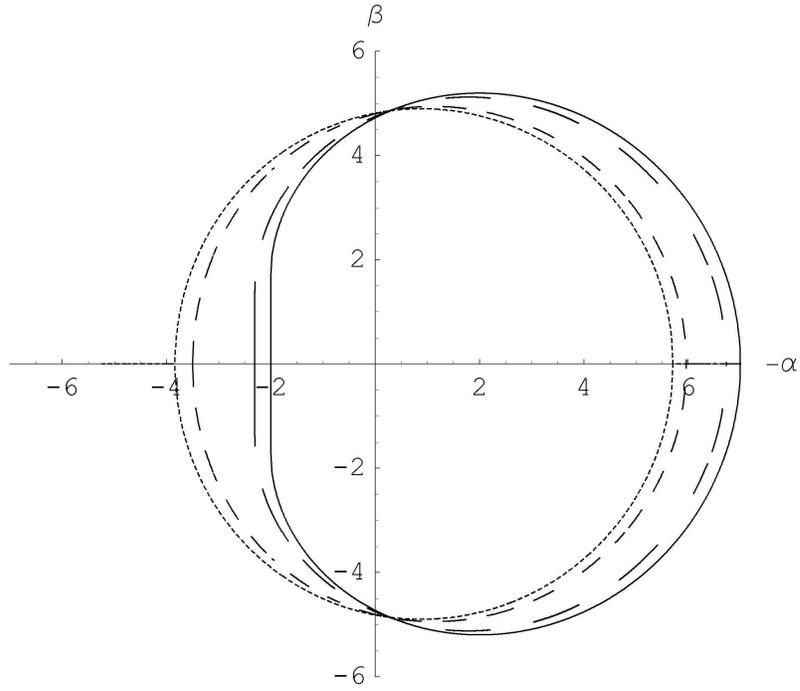}
\end{center}
\caption{Mirages around black hole for different angular positions
of a distant observer and  the spin $a=1$. Solid, long
 dashed, short dashed and dotted lines correspond to
$\theta_0=\pi/2, \pi/3, \pi/6$ and $\pi/8$, respectively.}
 \label{Fig2b}
\end{figure}

Let us consider different values for the angular positions of a
distant observer $\theta=\pi/2,\pi/3$ and $\pi/8$ for the spin
parameter $a=0.5$ (Fig. \ref{Fig2a}) and $\theta=\pi/2,\pi/3,
\pi/4$ and $ \pi/6$ for $a=1.$ (Fig. \ref{Fig2b}). From these
Figures one can see that angular positions of a distant observer
could be evaluated from the mirage shapes only for rapidly
rotating black holes ($a \sim 1$), but there are no chances to
evaluate the angles for slowly rotating black holes, because even
for $a=0.5$ the mirage shape differences are too small to be
distinguishable by observations. Indeed,  mirage shapes weakly
depend on the observer angle position for moderate black hole spin
parameters.

\section{Projected parameters of the space RADIOASTRON\\ interferometer}

During this decade the space radio telescope RADIOASTRON will be
launched. This project was initiated by Astro Space Center (ASC)
of Lebedev Physical Institute of Russian Academy of Sciences (RAS)
in collaboration with other institutions of RAS and RosAviaKosmos.
Scientists from 20 countries develop the scientific payload for
the satellite and will provide a ground base support of the
mission. The project was approved by RAS and RosAviaKosmos and is
smoothly developing. This space based 10-meter radio telescope
will be used for space--ground VLBI measurements. The
measurements will have extraordinary high angular resolutions,
namely about 1~-- 10 microarcseconds (in particular about 8
microarcseconds at the shortest wavelength 1.35 cm and a standard
orbit and could be about 0.9 microarcseconds for the high orbit at
the same wavelength. For observations four wave bands will be used
corresponding to $\lambda=1.35$~cm, $\lambda=6.2$~cm,
$\lambda=18$~cm, $\lambda=92$~cm.

An orbit for the  satellite was chosen with high apogee and with
period of satellite rotation around the Earth 9.5 days, which
evolves as a result of weak gravitational perturbations from the
Moon and the Sun. The perigee is in a band  from 10 to 70 thousand
kilometers, the apogee is a band from 310 to 390 thousand
kilometers. The basic orbit parameters will be the following: the
orbital period is p = 9.5 days, the semi-major axis is a = 189 000
km, the eccentricity is e = 0.853, the perigee is H = 29 000 km.

A detailed calculation of the high-apogee evolving orbit can be
done if the exact time of launch is known.

After several years of observations,  it would be possible to move
the spacecraft  to a much higher orbit (with apogee radius about
3.2 million km), by additional spacecraft maneuver using
gravitational force of the Moon. In this case it would be
necessary to use 64--70~m antennas for the spacecraft control,
synchronizations and
telemetry \footnote{http://www.asc.rssi.ru/radioastron/}.

 The fringe sizes
(in micro arc seconds) for the apogee of the above-mentioned orbit
and for all RADIOASTRON bands are given in Table 2.

\begin{table}[th!]
\begin{center}
\caption[]{The fringe sizes (in micro arc seconds) for the
standard and advanced apogees $B_{max}$ (350 000 and 3 200 000~km
correspondingly).}
\begin{tabular}{|c|c|c|c|c|}
\hline
$B_{max}({\rm km}) \backslash \lambda ({\rm cm})$ & 92 & 18 & 6.2 & 1.35 \\
\hline \hline
$3.5\times 10^{5}$ & 540 & 106 & 37 & 8 \\
\hline
$3.2 \times 10^{6}$ & 59 & 12 & 4 & 0.9 \\
\hline
\end{tabular}
\end{center}
\label{tabl1}
\end{table}

Thus, there are non-negligible chances to observe such mirages
around the black hole at the Galactic Center and in nearby AGNs
and microquasars in the radio-band using RADIOASTRON facilities.

We also mention that this high resolution in radio band will be
achieved also by Japanese VLBI project VERA (VLBI Exploration of
Radio Astrometry), since angular resolution will be at the 10~$\mu
as$ level \cite{Sawada00,Honma01,Honma02}. Therefore, there only a problem
to have a powerful radio source to illuminate a black hole to be
detectable by such radio VLBI telescopes like RADIOASTRON or VERA.

\section{Searches of mirages near Sgr $A^*$ with RADIOASTRON}

 Observations of Sgr A$^*$ in radio, near-infrared
and X-ray spectral bands develop very rapidly
\citep{Lo98,Lo99,Genzel03,Ghez04,Baganoff01,Bower02,Bower03,Narayan03,Bower04}\footnote{An
interesting idea to use radio pulsars to test a region near black
hole horizon was proposed in \cite{Pfahl03}.} also because it
harbours the closest massive black hole. The mass of this black
holes is estimated to be $4\times 10^6~M_{\odot}$
\citep{Bower04,Melia01,Ghez03,Schodel03} and its intrinsic size
from VLBA observations at wavelengths $\lambda=2$ cm, 1.3~cm,
0.6~cm and 0.3~cm  \citep{Bower04}.

Similarly to \cite{Falcke00} we propose to use VLBI technique to
observe the discussed mirages around black holes. They used
ray--tracing calculations to evaluate the shapes of shadows. The
boundaries of the shadows are black hole mirages (glories or
``faces'') analyzed earlier. We use the length parameter
$r_g=\dfrac{GM}{c^2}=6 \times 10^{11}$ cm to calculate all values
in these units as it was explained in the text. If we take into
account the distance towards the Galactic Center $D_{\rm
GC}=8$~kpc then the length $r_g$ corresponds to angular sizes
$\sim 5\mu{\rm as}$. Since the minimum arc size for the considered
mirages are about $2 r_g$, the standard RADIOASTRON resolution of
about $8~ \mu{\rm as}$ is comparable with the required precision.
The resolution in the case of the higher orbit and shortest
wavelength is  $\sim 1\mu{\rm as}$ (Table~2) good enough to
reconstruct the shapes. Therefore, in principle it will be
possible to evaluate $a$ and $\theta$ parameters after mirage
shape reconstructions from observational data even if we will
observe only the bright part of the image (the bright arc)
corresponding to positive parameters $\alpha$. However,
\citet{Gammie04} showed that black hole spin is usually
 not very small and could reach 0.7--0.9 (numerical simulations of relativistic
 magnetohydrodynamic flows give $a \sim 0.9$).
Taking into account detections of 106 day cycle in Sgr A$^*$ radio
variability seen at 1.3 cm and 2.0~cm by \cite{Zhao01} at Very
Large Array (VLA), \cite{Liu02} suggested a procedure to evaluate
the black hole spin assuming that the variability could be caused
by spin induced disk precession.
  
  Moreover, the recent analysis  by \citet{Aschenbach04}
 of periodicity of X-ray flares from the Galactic Center black
 hole gives an estimate for the spin as high as
 $a=0.9939^{+0.0026}_{-0.0074}$.
Actually, the authors used generalizations of the idea proposed by
\citet{Melia01a} that the minimum rotation period for
Schwarzschild black hole (for an assumed black hole mass of $2.6
\times 10^6M_\odot$) is about $P_0 \approx 20$ minutes and could
be in the range $P_0 \in [2.6,36]$ minutes depending on the black
hole spin and prograde and retrograde accretion flows generating
the quasi-periodic oscillations. Using this idea and analyzing
quasi-periodic variabilities in a infrared band \citet{Genzel03}
concluded that the black hole spin should be $a \sim 0.5$.
However, this conclusion is based on the assumption that the
emitting region is located at the marginally stable orbit,
therefore if the periodicity is related to the emitting gas motion
around the black hole, we should conclude that the black hole spin
is $ a \gtrsim 0.5$. One could also mention that such a
determination of the black hole spin is indirect and actual
typical frequencies for real accretion
 flows could be rather different from frequencies considered by the
 authors. We may summarize by saying that there are indications
 that the spin of the Galactic Center black hole can be very high,
 although this problem is not completely solved up to date.

\section{Discussion}

As stated in Section 2, the part  of Kerr quasi--rings formed by
co-rotating photons is much brighter with respect to the opposite
side (i.e. the part of the image formed by counter--rotating
photons) and in principle can be detected  much more easily.
However, even the bright part of the quasi--ring  can give
information about mass, rotation parameter and inclination angle
of the black hole. Of course, if the black hole--observer
distance is unknown, the black hole mass can be evaluated in units
of the distance. Even if the faint part of image (which is formed
by counter--rotating photons) is not detectable, one can try to
reconstruct the shape of the total image searching for the best
fit of the full image using only the bright part of the image.

Of course, we have such superpower laser (greater than Gigawatt)
which \citet{Holz02} wanted to use to investigate a black hole by
an active way and, in principle, it is possible to infer the spin
parameter and inclination angle by analyzing of mirage shapes
formed by retro-photons.

\citet{Holz02} mentioned that the structure of such retro-Macho
images will be unresolvable, because of even for very close black
holes (say at the age of the solar system), the angular extent of
retro-Macho images remain less than a milliarcsecond. If we we
could assume that this black hole could be at the edge of the
solar system. We could not restrict ourselves to speculate on
future observational facilities. For example, future Space
Interferometry Mission
(SIM)\footnote{http://planetquest.jpl.nasa.gov/SIM/}, GAIA
satellite~\footnote{http://astro.estec.esa.nl/gaia} and
FAME \footnote{The FAME Concept study. Report by~\citet{John99} is
available at http://www.usno.navy.mil/FAME/publications, see also
\cite{Salim02}.}  could resolve such images, because they will
have a sufficient angular resolution for our example (for example,
over a narrow field of view SIM could achieve an accuracy of 
1~$\mu as$, a similar accuracy will have GAIA), but unfortunately
its limiting magnitude will be about 20~mag (the limiting
magnitude for GAIA $V \sim 20$), and since the limiting distance
to discover such phenomenon is restricted by \citet{Holz02}
\begin{equation}
D_{L} =0.02 {\rm pc} \times
\left[10^{(m-30)}/2.5(M/M_\odot)^2\right]^{1/3},
          \label{con_1}
\end{equation}
thus, for SIM limiting magnitude and even for massive black hole
($M=10M_\odot$), we obtain that such a  retro-Macho should be
inside of solar system since $D_{L}\sim 10^{-4}$pc. Therefore, to
discover such phenomenon we need instrument like SIM concerning
angular resolution, but limiting magnitude should on 10 mag better
than SIM. In this case there is a possibility not only to observe
such retro-Macho at the the edge of the solar system but also to
determine its angular momentum.

\section{Conclusions}

We could summarize that angular resolution of the space
RADIOASTRON interferometer will be high enough to resolve radio
images around black holes therefore analyzing the shapes of the
images  one could evaluate the mass  and the spin $a$ for the Kerr
black hole inside the Galactic Center and a position angle
$\theta_0$ for a distant observer and as it is clear a position
angle could be determined by more simple way for rapidly rotating
black holes $a \sim 1$ (in principle, measuring the mirage shapes
we could evaluate mass, inclination angle and spin parameter if we
know the distance toward the observed black hole. Otherwise one
can only evaluate the spin parameter in units of the black hole
mass since even for not very small spin $a=0.5$ we have very weak
dependence on $\theta_0$ angle for mirage shapes and hardly ever
one could determine $\theta_0$ angle from the mirage shape
analysis. Moreover, we have a chance to evaluate parameters $a$
and $\theta$ (for rapidly rotating black holes) if we reconstruct
only bright part of the mirages (bright arcs) corresponding to
co-moving photons $(\alpha>0)$. However, for slow rotating black
holes $\alpha \lesssim 0.5$ it would be difficult to evaluate
parameters $a$ and $\theta$ because we have very slow dependence
of mirage shapes on these parameters.

However, there are two kind of difficulties to measure mirage
shapes around black holes. First, the luminosity of these images
or their parts (arcs) may not be sufficient to being detectable by
RADIOASTRON. However, numerical simulations by
\citet{Falcke00,Melia01} give hope that the luminosity could be
not too small at least for arcs of images formed by co-rotating
photons $(\alpha>0)$. Second, turbulent plasma could give
essential broadening of observed images \cite{Bower04}, the
longest interferometer baseline $b_{max} \sim 350000$~km (or for
higher orbit $b_{max}\sim 3.2 \times 10^6$~km) and for this case
we have similar to \cite{Bower04} length scale in the scattering
medium is $l=(D_{scattering}/D_{GC})\times b_{max} \sim
4.4*10^3$~km (or $l=4.4 \times 10^4$~km for the higher orbit).
Thus, the scale could be less or more than the predicted and
measured values of the inner scale, which are in the range $10^2$
to $10^{5.5}$~km \citep{Wilkinson94,Desai01,Bower04}, thus the
broadening the images could be essential but it is not very easy
to calculate it in details for such parameters.

Recent observations of simultaneous X-ray and radio flares at 3
mm, 7 mm, 1.3 cm and 2 cm with the few-hundred second rise/fall
timescales gave indirect evidences that X-ray and radio radiation
from the close vicinity of Sgr A$^*$ was detected because of that
is the most natural interpretation of these flares. However,
another interpretations of these flares could not be ruled out and
in this case an optical depth for radio waves at 1.3 cm wavelength
toward  Sgr A$^*$ may be not very small.

Few years ago a possibility to get images of nearby black holes in
X-ray band was discussed by \citet{White00,Cash00}, moreover
\citet{Cash00} presented a laboratory demonstration of the X-ray
interferometer. If the project will be realized, one could get
X-ray images of black holes with $0.1 \times 10^{-6}$ arcsec
resolution, thus using this tool one could detect X-ray images
around the Galactic Centre and around the black hole in M87
Galaxy.

One could mention also that if the emitting region has a
degenerate position with respect to the line of sight (for
example, the inclination angle of an accretion disk is $\gtrsim
85^0$) strong bending effects found by \citet{Matt93} and analyzed
later by \citet{zak_rep03b} do appear. In this case, the mirage
shapes will be strongly distorted  \citep{Beckwith04}.

In spite of the difficulties of measuring the shapes of images
near black holes is so attractive challenge to look at the ``faces''
of black holes because namely the mirages outline the ``faces'' and
correspond to fully general relativistic description of a region
near black hole horizon without any assumption about a specific
model for astrophysical processes around black holes (of course we
assume that there are sources illuminating black hole
surroundings). No doubt that the rapid growth of observational
facilities will give a chance to measure the mirage shapes using
not only RADIOASTRON facilities but using also other instruments
and spectral bands (for example, X-ray interferometer
\cite{White00}).

 Talks on the subject were
presented by AFZ at conferences on General Relativity SIGRAV-2004
(Salerno, Italy, see contribution \cite{Zak_SIGRAV04}) and
ERE-2004 (Madrid, Spain) and authors thank participants for
constructive discussion. We also thank profs. J.~Miller and
L.~Rezzolla for fruitful comments after a talk on the subject
presented by AFZ at SISSA.

\section*{Acknowledgements}

 AFZ would
like to thank Dipartimento di Fisica Universita di Lecce and INFN,
Sezione di Lecce  for a hospitality (the work was finished in
Lecce). AFZ expresses also his acknowledgement to organizers of
the XXVII International Workshop on Fundamental Problems of High
Energy Physics and Field Theory ``Black holes on Earth and Space'',
especially to Academician A.A. Logunov for the kind invitation to
present the invited talk at the Workshop.


\baselineskip = 0.98\baselineskip

\end{document}